# Double tips for in-plane polarized near-field microscopy and spectroscopy

Patryk Kusch[1*], José Pareja Arcos[1], Aleksei Tsarapkin[2], Victor Deinhart[2], Karsten Harbauer[3], Katja Höflich[2] and Stephanie Reich[1]

[1] Freie Universität Berlin, Fachbereich Physik, Berlin, Germany
[2] Ferdinand-Braun-Institut, Leibniz-Institut fuer Hoechstfrequenztechnik (FBH), Berlin Germeny
[3] Institute for Solar Fuels, Helmholtz-Zentrum Berlin fuer Materialien und Energie GmbH

*corresponding author: patryk.kusch@fu-berlin.de

ABSTRACT

*Near-field optical microscopy and spectroscopy provide high-resolution imaging below the diffraction limit, crucial in physics, chemistry, and biology for studying molecules, nanoparticles, and viruses. These techniques use a sharp metallic tip of an atomic force microscope (AFM) to enhance incoming and scattered light by excited near-fields at the tip apex leading to high sensitivity and a spatial resolution of a few nanometers. However, this restricts the near-field orientation to out-of-plane polarization, limiting optical polarization choices. We introduce double tips that offer in-plane polarization for enhanced imaging and spectroscopy. These double tips provide superior enhancement over single tips, although with a slightly lower spatial resolution (~30nm). They enable advanced studies of nanotubes, graphene defects, and transition metal dichalcogenides, benefiting from polarization control. The double tips allow varied polarization in tip-enhanced Raman scattering and selective excitation of transverse-electric and -magnetic polaritons, expanding the range of nanoscale samples that can be studied.*

Near-field optical microscopy and spectroscopy exploit localized plasmonic near fields to overcome the diffraction limit of light[1]. These powerful nanoimaging tools include scattering-type scanning near-field optical microscopy (s-SNOM), tip-enhanced photoluminescence (TEPL), and tip-enhanced Raman scattering (TERS). Scattering-type SNOM is known for its manifold of impressive near-field images, highlighting, e.g., changes in the dielectric function with position and the distribution of doping and strain across a surface. It makes the propagation of surface polaritons directly visible and thereby gives access to the dispersion and losses of the hybridized light-matter excitations [2-6]. TERS provides chemical fingerprints in nanoscale processes to study chemical reactions at the molecular level and visualize individual molecules with nanometer resolution[7-20]. The position and distance between localized emitters are measured by TEPL with the potential for ultra-high resolution bioimaging[21].

The nanoscale resolution of near-field techniques relies on localized electric fields, that provide enhanced field intensities and high momenta. Both can significantly increase the excitation probability of dipole-carrying matter excitations [22]. At their core, the localized fields involve the dynamical coupling of electrons in collective motion (i.e. plasmons) with photons [23, 24]. The corresponding quasiparticle is named plasmon-polariton and can be exploited for manipulating light at subwavelength scales, confining electromagnetic fields to nanoscale volumes, and enhancing light absorption and scattering [25]. In near-field microscopy and spectroscopy sharp metal tips are brought into the proximity of a sample, typically placed within a few nanometers. Plasmon polaritons are excited at the tip by a focused laser beam, creating highly confined electromagnetic near-fields between the tip and the sample. The incoming, elastically (s-SNOM) or inelastically (TERS) scattered light of the probe or its luminescence (TEPL) can be amplified by the near-fieldes [8, 9]. The tip is scanned across the sample to detect the enhanced light as a function of tip position, resulting in an optical image with a resolution that is given by the size of the tip apex.

The enhancement in near-field techniques depends critically on the coupling between the localized field and the sample. The key parameters are the tip-sample distance and the polarization of the plasmonic near-field. In the commonly employed AFM setups, the tip-generated fields are mainly polarized out-of-plane along the *z-axis* (cf. Fig. 1A), whereas the dipole transition moments of excitations in two-dimensional materials, large organic molecules, and planar nanostructures are predominantly in-plane [26]. This misalignment limits tip-sample coupling and thus the tip-enhancement factors [26]. Controlling the near-filed orientation in scattering-type SNOM would also allow to selectively excite transverse magnetic (TM) and electric (TE) polaritons and waveguided modes, which has been highly challenging for out-of-plane polarization [27, 28].

Here we present a metallic double tip for the in-plane excitation in near-field microscopy and spectroscopy. The tip effectively implements scanning with a nanodimer, where the near-field is excited between two metal nanoparticles resulting in in-plane polarization [29, 30]. The double tip comprises two single tips that point towards each other and are grown by direct electron beam writing. As in the nanodimer case, their optical near-fields couple efficiently to in-plane polarized excitations in absorption, light scattering, and luminescence and allows to exploit optical polarization selection rules. The improved enhancement and polarization control is achieved with only a negligible loss in resolution due to the larger extension of the plasmonic hotspot, as we show in a series of near-field microscopy and spectroscopy experiments on graphene, transition metal dichalcogenides (TMDs), and boron nitride nanotubes (BNNTs) supported by finite difference time domain (FDTD).

RESULTS AND DISCUSSION

The excitation with in-plane polarization is realized with two metallic tips with a gap that is much smaller than the tip diameter, Fig. 1B. The coupling of the localized surface plasmons of both tips produces a near-field hotspot between them. When illumined with light of 480nm wavelength polarized parallel to the dimer axis, Fig 1D, a silver dimer is predicted to show a stronger hotspot than a single antenna under out-of-plane polarization. Crucially, the double tip hotspot is polarized mainly in-plane, whereas the single tip yields out-of-plane polarization. While structured AFM tips were reported before to increase field enhancement [31-35], their ability for polarization control has not been realized so far. The proposed silver nanodimer shows a plasmon resonance in the visible 420-500 nm, which is ideal for TERS and TEPL as they normally operate in the visible and near-IR spectral range.

Double-tips on AFM cantilevers as depicted in Fig. 1E were fabricated by focused ion and electron beam processing followed by glancing angle deposition of silver. The focused Ga ion beam was used to mill a well-defined plateau on a conventional AFM cantilever. Subsequently, a focused electron beam was used to locally dissociate a gaseous precursor, a process called focused electron beam-induced deposition (FEBID) [36-38]. The beam path during the direct writing process was carefully optimized for mechanical stability of the antennas and to realize gap sizes of about 50-70 nm that account for the expected thickness of the added silver layer. The final silver-coated dimer antennas had a total length of 5 μm and a gap size of about 50 nm which determines the spatial resolution and field enhancement (cf. methods in SI).

In the following, we demonstrate double tips for near-field imaging and TERS using in-plane polarization. Techniques like TERS and s-SNOM were used to evaluate the performance of the d tips by imaging topographies of 2D samples and nanotubes. Resolution and enhancement were compared between double and single silver tips using the dual s-SNOM operating in tapping mode.

We confirmed the near-field presence between the tip and substrate by recording approach curves at higher harmonics in s-SNOM. Adjusting the parabolic mirror for maximum signal at the 4$^{th}$ harmonic frequency, and when retracting the tip, we observed a strong signal decay within the first 100 nm, indicating the near-field's extension of some tens of nanometers.

Approach curves for both double and single silver tips showed expected behavior. The 4th harmonic near-field amplitude decreases rapidly within 100 nm tip-sample distance, highlighting that a near-field is generated[39], and the demodulation on the fourth harmonic efficiently suppresses the optical background signal.

Despite its two-tip character the double tips work for obtaining topography images through the AFM with some decrease in spatial resolution. Scattering-type SNOM and TERS setups deliver AFM topography images simultaneously with the optical microscopy and spectroscopy response. In fact, a reliable AFM system is a key requirement for demodulation in s-SNOM setups [40]. To demonstrate that the double tips are suitable for recording the nanoscale topography, we imaged an $MoS_2$ flake and a boron nitride nanotube (BNNT). The edge of the $MoS_2$ flake is well observable when recorded with the conventional cantilever, Fig. 2A, and the double tip, Fig. 2B. Both images show a second step inside the flake and deliver similar sample height (double: 61 nm, single: 60 nm). We estimate the spatial resolution by scanning over the flake edge, Fig. 2E. We estimate the spatial resolution to be 30 nm for the double tip, which is decreased compared to the 5 nm determined for the conventional tip.

With a very narrow nanoscale structure, we can directly observe the splitting of AFM features due to the double tip. The topography of BNNTs recorded with a single tip, Fig. 2C, shows a nanotube bundle of 8 nm in height that gets thinner towards the end. At the top of the image, two individual BNNTs are visible. The same topography taken with the double tips, Fig. 2D, excellently reproduces the BNNT bundle. The single BNNTs, however, exhibit a double structure. This is clearly shown as two maxima in the height profile of the double tip, blue line in Fig.2F, whereas the conventional tip shows only a single peak (red). Each of the double tips provides the topography information that is averaged in the image. Since the tip spacing (30 nm) is wider than the diameter of the tip apex, we observe the BNNTs twice, an artifact that also occurs with broken AFM tips. However, the optical near-field of the double tips has a single hotspot, Fig. 1, so the double image is absent in optical microscopy. The 4th harmonic near-field optical image in Fig. 4G was recorded simultaneously with the topography, Fig. 3D; it is free of artifacts and shows the BNNTs without any double structure.

Double tips allow for finely tuning the polarization direction in scattering-type SNOM by controlling the laser polarization, which has been elusive so far. Two-dimensional flakes support ordinary, transverse electric (TE) and extra-ordinary, transverse magnetic (TM) waveguided modes. These waveguided modes couple to the optically active excitations[27, 41]. In the strong and ultra-strong coupling regime, the materials excitations become of mixed light-matter or polaritonic character, which is sometimes labeled as self-hybridization to distinguish it from the coupling with photons of an external cavity[42]. Single-tip experiments predominantly image TM-derived polaritons that have a strong polarization component normal to the plane [41, 43]. Transverse-electric modes may also be excited in this configuration, but the much stronger TM modes overlap and often mask the TE polaritons [44]. On the other hand, only TE modes may be excited with in-plane polarized near fields as provided by double tips.

The dispersions of TE and TM-derived polaritons differ so that polaritons excited with in-plane and out-of-plane polarized near-fields are expected to have different wavelengths at the same frequency[27, 41]. We measured the s-SNOM near-field images of propagating exciton polaritons in an $MoS_2$ flake (65 nm in height). Figure 3A shows the exciton-polaritons excited with in-plane polarization by the double tip, whereas Fig. 3B is the image recorded under out-of-plane single-tip polarization. The line scans in Fig. 3C and D demonstrate the expected difference in polariton wavelength [27, 41]: The single tip excites the TM mode of the $MoS_2$ flake with $\lambda$ = 235 nm at 2.54 eV excitation energy, Fig. 3A, whereas we observe the TE mode under double-tip excitation with a shorter wavelength ($\lambda$ = 170 nm) or longer wavevector, Fig. 3B. Double tips may be used to selectively switch from TE to TM modes by changing the near-field polarization. Previous observations of TE polaritons in two-dimensional materials required flakes with ~100 nm thickness[27]; double tips will be able to excite the transverse electric excitations down to the thinnest flakes.

In addition to near-field optical microscopy, the dual s-SNOM offers the opportunity to record tip-enhanced PL spectra. We record TEPL spectra as a function of tip position along a line perpendicular to the edge of a $WSe_2$ monolayer, Fig. 5. The topography and optical near-field image, Fig. 4A and B, show bubbles in the monolayer that were produced during the exfoliation and allow identifying the edge of the $WSe_2$ flake. TEPL spectra taken on the monolayer, blue circle in Fig. %A, show a 30% increase in luminescence intensity compared to when the tip is on the gold substrate, red circle, see blue and red spectra in Fig. 4C. A TEPL scan across the monolayer edge, white line in Fig. 4B, shows the drop in TEPL intensity when the double tips move through the edge, Fig. 4D. The step-like function allows extracting the spatial resolution of 38 nm. For comparison, we measured the enhancement by a single tip in gap mode, Fig. 4E, and found it to be similar to the double tip. The photoluminescence of $WSe_2$ can be excited with in-plane and out-of-plane polarized light. The out-of-plane polarization (single tip) is less efficient in the far field, but the inferior polarization configuration is counteracted by the higher enhancement obtained in the gap mode configuration of a single tip on top of a metal substrate resulting in comparable intensities for the double and single tips. The single tip has a higher spatial resolution (~ 20 nm, Fig. 4E) and, indeed, the gap mode configuration may yield a resolution of 5 nm and less under ideal conditions.

Another near-field technique that greatly benefits from polarization control is tip-enhanced Raman scattering that has strong polarization dependent selection rules[45-47]. We use graphene to probe the performance of the double tips in TERS because the Raman-active modes of graphene have a constant cross-section and require in-plane polarization of the incoming and scattered light. Many TERS studies have been reported for graphene allowing to benchmark the performance of the double tips[18, 46, 47]. We record a topography and a near-field optical image of a graphene sheet, Fig. 5A and B, with the dual s-SNOM after which TERS is performed. In both images, the graphene edge is clearly observed including features like bubbles and wrinkles. To determine the TERS enhancement created by the double tips, we recorded TERS spectra of the 2D mode when the tip is close to the edge, spot 1 in Fig. 5A, and 30 nm away from the edge, spot 2. The integrated 2D intensity decreases when moving the tip 30 nm away from the graphene, Fig. 5C blue and red spectra, indicating resolution and enhancement of the double tip.

It is challenging to quantify TERS enhancement in measurements on 2D materials, because the Raman spectra tend to be dominated by the far-field Raman response[46]. To estimate the enhancement factor *M* we use the relation

$$M = \left[\frac{I_{NF} - I_{FF}}{I_{FF}}\right] \times \frac{A_{FF}}{A_{NF}},$$

where $I_{NF}$ is the integrated TERS and $I_{FF}$ the integrated Raman intensity away from the tip, $A_{NF} = 8 \cdot 10^2$ nm² the area under the tip and $A_{FF} = 2 \cdot 10^5$ nm² the area of the laser spot [31]. By recording the 2D intensity when the tip is on graphene to the intensity measured with the tip on the bare substrate, Fig. 5C spectra @1 (blue) and @2 (red), we find $I_{NF}/I_{FF} = 3.8$. This results in an enhancement factor *M*~700, an excellent value for TERS experiments. The enhancement for the single tip in out-off plane configuration is 60, i.e., ten times smaller than the double tip. We note that single TERS tips, especially custom-designed pyramid tips or in gap mode configuration, can also achieve TERS enhancement factors on the order of $10^2$ [48-50], but this required careful optimization in contrast to the proof of concept experiments performed here.

A key application of TERS is chemical-sensitive imaging of surfaces and nanostructures with sub-wavelength resolution[51]. To demonstrate the resolution of the double tips in enhanced light scattering, we scan it across a graphene edge while recoding tip-enhanced Raman spectra, Fig. 5D. Since the enhancement of the double tips (blue symbols) is higher than for the single tip (grey), the edge is much better detectable with the double tips. From the measured intensities, we extract a resolution of 32 nm for the double tips, blue line fit in Fig. 5D, which agrees with the topography scan taken with these tips in Fig. 5A and the resolution obtained by TEPL in Fig. 4D. The single silver tip has a resolution of less than 20 nm, Fig. 5D gray line, but the sensitivity is much lower and the position of the edge is much harder to observe with chemical sensitivity.

To record TERS from a quasi one-dimensional structure with the double tips we measure the intensity of the D band around the edge, Fig 5F. The D mode is activated by defects in graphene, which may be a point defect or a line like an edge [52-54]. When the double tip is positioned directly at the graphene edge, we observe a strong D band, Fig 5E top spectrum. When retracting the tip or moving it only 20 nm away from the edge the mode disappears, Fig. 5E middle and bottom. This highlights the exceptional ability of the double tips to locally excite in-plane polarized Raman processes. The edge of graphene becomes visible as a quasi-one-dimensional defect structure. For reference, we attempted to measure the D mode with a single tip and far-field Raman scattering, but their sensitivity was insufficient to detect any edge-induced scattering.

In conclusion, we proposed to use AFM double tips to provide in-plane polarized excitation in near-field optical microscopy and tip-enhanced spectroscopy (TERS and TEPL). The double tips were grown free-standing using focused electron beam induced deposition onto conventional AFM cantilevers. We showed that SNOM images taken with such tips allow to control the polarization during excitation leading, for example, to the selective excitation of TE and TM waveguided modes in 2D flakes of transition metal dichalcogenides. The double tip provided an order of magnitude higher enhancement of the cross-section in tip-enhanced Raman scattering on a graphene monolayer, which can be further optimized with smaller gaps and tailored tip design. The increase is due to the orientation of the exciting near field and a better coupling of the double tips to the induced Raman dipole. It comes at the cost of a lower resolution that roughly dropped by a factor of two when comparing single and double tips. The in-plane excitation in near-field microscopy and spectroscopy offers a powerful option for scattering-type SNOM, TERS, and TEPL leading to new insights into light-matter interaction and to using optical selection rules in near-field experiments in future.

ACKNOWLEDGEMENTS

We thank Niclas S. Mueller for helpful discussions, exchange of ideas, and support of the near-field simulation. The focused electron and ion beam processing was performed in the CoreLab

Correlative Microscopy and Spectroscopy at Helmholtz-Zentrum Berlin. The glancing angle deposition equipment is part of the Helmholtz Energy Materials Foundry (HEMF, GZ 714-48172-21/1), which was funded by the German Helmholtz Association. AT, VD and KH acknowledge financial support from DFG under grant HO 5461/3−1. Parts of the work were supported by the European Research Council ERC under grant DarkSERS (772108). PK acknowledge the DFG for funding (KU4034 2-1). This work was supported by the SupraFAB Research Facility and the Focus Area NanoScale at Freie Universität Berlin.

**Supporting Information**
METHODS

**Near-field simulations**: Silver spherical nanoparticle and nanoparticle dimers were simulated using the finite difference time domain method as implemented in Lumerical FDTD Solutions from Ansys. We use silver nanoparticles with a diameter of $d = 50$ nm, which was identified as the representative diameter of the tips produced in this work. Silver was described with the dielectric constant measured by Yang et al[55]. We use a mesh override region of 0.5 nm in all directions across the nanostructures. The nanostructures were illuminated from the top with a total-field scattered field source with the electric field polarized along the dimer axis. The electric field was recorded with an electric field monitor. The field enhancement was calculated as the absolute value of the local electric field normalized by the incident field[55].

**Fabrication of the double tips:** Double tips were produced by electron beam-induced deposition and covered with a thin silver film by glancing angle deposition[36, 56]. The fabrication was performed in a Zeiss Crossbeam 340 dual beam instrument and started from the commercial AFM cantilevers Arrow NCPT. Using focused Ga ions of 30 keV energy and a beam current of 1.5 nA a well-defined flat plateau was milled on the tip. For the following deposition step an electron beam of 15 keV energy and a beam current of about 250 pA was used to locally decompose Pt($\eta^5$-CpMe)Me$_3$ as the precursor compound. The optimum compromise between mechanical stability and spatial resolution was achieved for a relatively broad antenna radius at the base that decreases towards a tip of very small radius. Therefore, the beam was scanned in spiraling path starting with a base radius of 250 nm that linearly decreased and ends in a single point. The spiral pitch defined the displacement of two consecutive spiral windings resulting a total horizontal length in projection of 2 µm. To account for the decreasing deposition rate with structure height (which is a consequence of the diffusion-driven precursor supply) the spiral pitch was decreased along the base towards the tip according to a power law with an exponent of 0.5. The beam path was rastered with a constant pixel distance of 0.1 nm and constant dwell time of 200 µs. To realize a quasi-parallel writing with the beam jumping back and forth between the two arms of the dimer the two spiral patterns for the antenna arms were split into 10 equal parts. The gap width in the entire dimer pattern is 125 nm resulting in about 50 - 70 nm gap widths in the deposited structure. The total length of the obtained dimer antenna is about 4 µm with an angle between the antenna arms and the surface of 50°. The optimized pattern is made available as example part of the patterning toolbox FIBomat[57].

The fabricated dimer antennas on the AFM cantilevers were then covered with silver by electron beam evaporation (Scia Vario 100) under glancing angle conditions with a sample tilt of 85° and a rotation speed of 20 rpm and a base pressure of 3.2E-7 mbar. Using a beam current of 180 mA a target thickness of 50 nm was deposited at an evaporation rate of 0.6 nm/s which resulted in about 20 nm of silver layer thickness on the antenna structures.

**Near-field experiments:** The TERS and s-SNOM experiments were carried out by the dual s-SNOM[4, 19, 58]. It is based on a conventional neaSNOM (neaSPEC) that is extended with a spectrometer to detect inelastically scattered light and luminescence. As an excitation source, we use a c-Wave laser (Hübner Photonics) that guarantees wavelength tunable cw excitation in the range 450-650 nm and 900-1300 nm. The setup operates with side illumination, where the laser is focused onto the tip apex by a parabolic mirror at an angle of 30° with respect to the sample. The polarization of the laser in a single tip measurement is parallel to the tip apex (*p*-polarized). The focused light excites plasmons that generate a strongly localized *z*-polarized near-field, i.e., out-of-plane. This near-field enhances both the incoming and emitted light (elastic and inelastic scattering, luminescence). For near-field imaging with the s-SNOM, the elastically backscattered light is detected by a single line silicon CCD. The inelastically scattered light is guided to a spectrometer equipped with a silicon detector (iQuos, Andor) enabling TERS and TEPL. In the SNOM mode, the AFM operates in tapping mode leading to a modulation of the near-field signal. By demodulating the recorded signal on higher harmonics (here the third and fourth harmonic) and by applying the pseudo heterodyne detection scheme we suppress the far-field background in the signal. The recorded signal is split into amplitude and phase, granting access to the absorption and reflection of the near-field optical signal with a resolution down to 20 nm[59]. The tip tapping frequency is 300 kHz and the amplitude 40 nm. To excite a near-field at the double tip we change the polarization of the laser to *s*-polarized (perpendicular to the tip axis) using a half-wave plate. In this configuration the near-field points from one tip to the other and is oriented parallel to the sample.

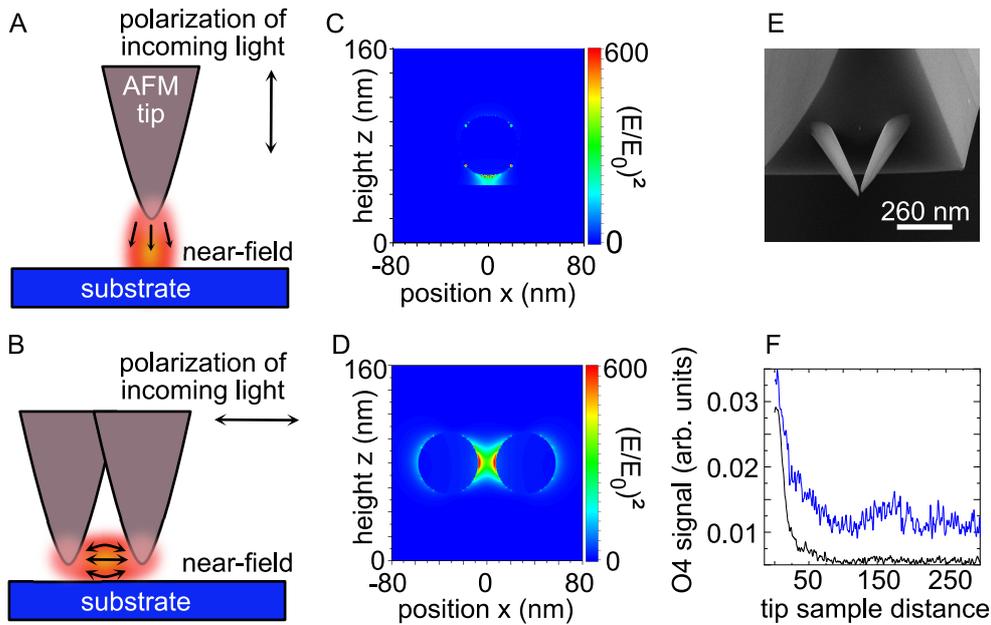

*Figure 1* **A** *Scheme of a conventional single tip used in TERS and s-SNOM. The polarization of the incident light is p-polarized as indicated. The orientation of the near-field is mainly out-off plane, perpendicular to the substrate.* **B** *Scheme of a double tip used in this work.* **C** *Calculated near-field between a single silver particle and a gold substrate (gap-mode configuration).* **D** *Calculated near-field between two silver nanoparticles that mimic the apex of the double tip. The orientation of the near-field is mainly in-plane, parallel to the substrate* **E** *SEM image of a double tip grown by FEBID.* **F** *Amplitude as a function of tip-sample distance demodulated at the fourth harmonic recorded at 480nm excitation with the double tip (blue, s polarized), and a single silver tip (black, p polarized). Both curves show the expected decay of the amplitude signal within 50 nm as is characteristic for near-fields.*

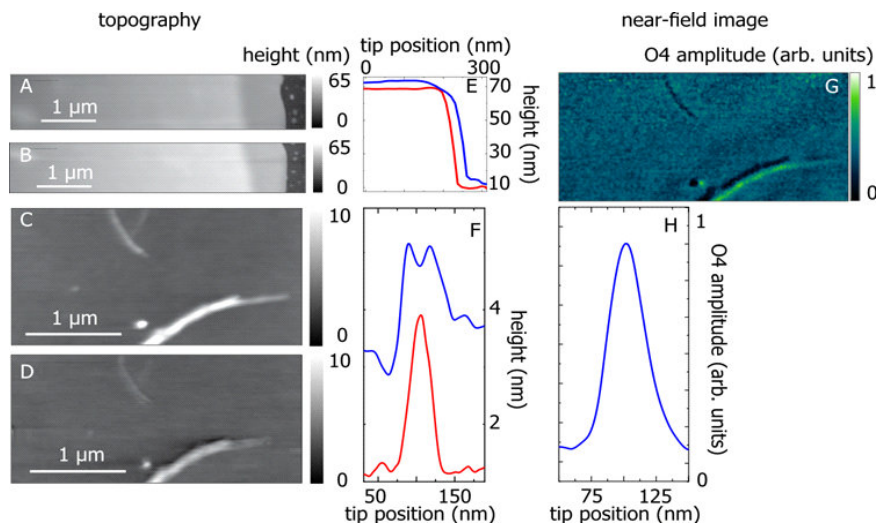

*Figure 2 Topography of an MoS$_2$ flake taken with* **A** *a single and* **B** *a double tip; topography of BNNTs on silicon recorded with* **C** *a single and* **D** *double tip.* **E** *Cross section extracted from* **A** *and* **B** *showing the height of the MoS$_2$ flake as a function of tip position, red: single tip, blue: double tip.* **F** *Cross section of the BNNTs for both tips, same colors as in* **E**. **G** *Optical amplitude image of BNNTs recorded with a double tip in the s-SNOM.* **H** *Cross section extracted from* **G** *at the same position as the topography cross section.*

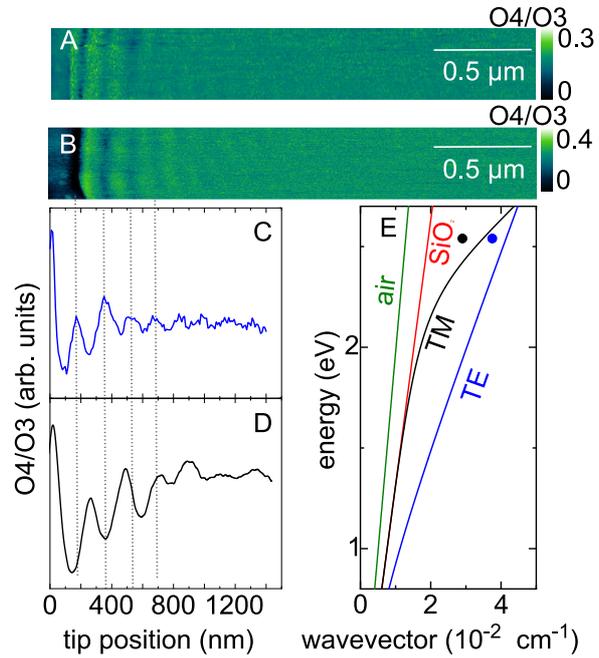

*Figure 3 Optical near-field images of few-layer MoS$_2$ with **A** a double tip and **B** a single silver tip. **C** and **D** extracted amplitude profiles from panels **A** and **B**, respectively. **E** predicted dispersion for TE and TM waveguided modes in MoS$_2$ flakes. The blue and black dots are determined from the near-field images in panels **A**, **B**. For completeness the light dispersion in air and SiO$_2$ are shown.*

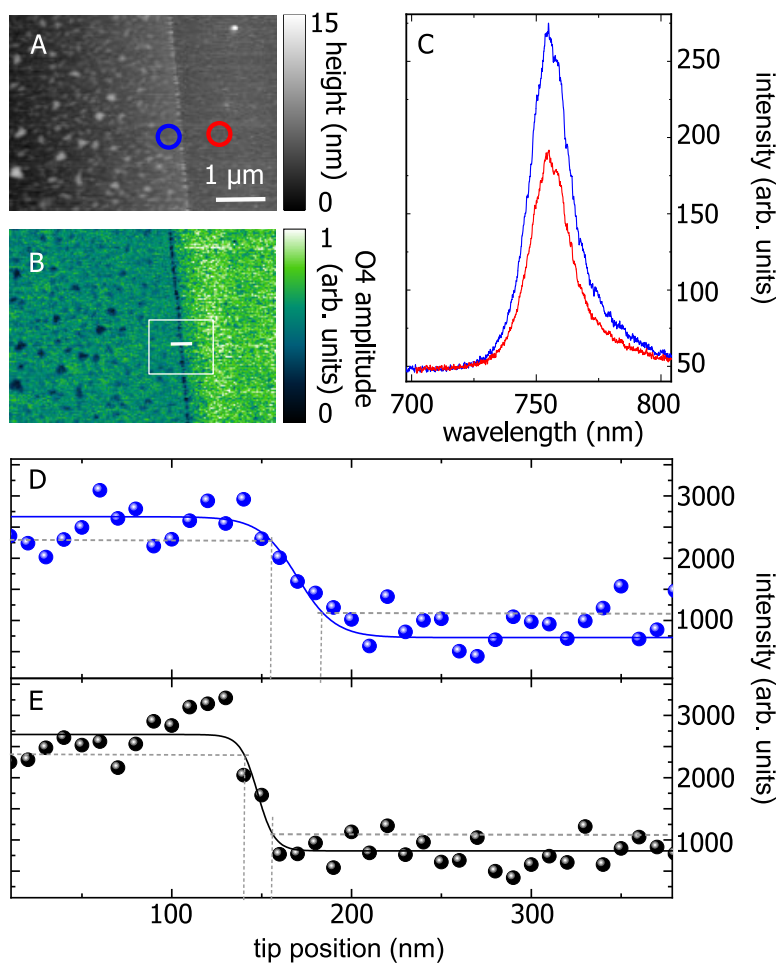

Figure 4 **A** Topography of a WSe$_2$ flake on a Au substrate. **B** Corresponding optical near-field image. **C** TEPL spectra recorded when the tip was located on WSe$_2$ (blue) and on the Au substrate (red). **D** PL intensity as a function of tip position recorded while the double tip was scanned perpendicular to the edge (white line in **B**). **E** Same as in **D**, but with a single tip.

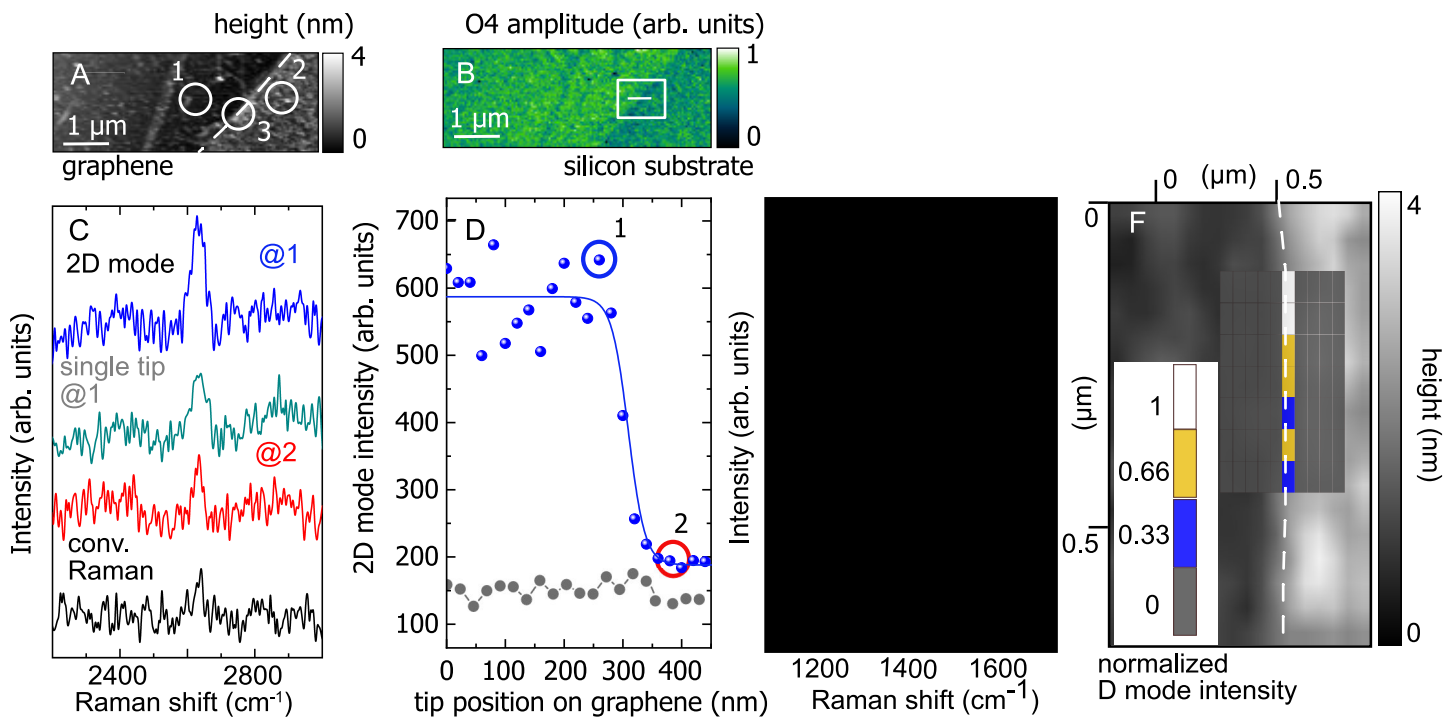

*Figure 5* **A** Topography and **B** near-field image of a graphene flake on Si recorded with the double tip. **C** Raman spectra of the 2D mode of graphene recorded at two different positions on the sample with the double tip (blue and red, as indicated in **A**). TERS spectra recorded with a single tip (grey) and without tip (black) are shown for comparison. **D** Raman intensity of the 2D band as a function of tip position for the double (blue) and single (grey) tip. The tips were moved along a line perpendicular to the graphene edge. **E** Raman spectra taken on the substrate (bottom black, position 2 in A), the graphene flake (middle, blue @1), and the graphene edge (top grey, @3). The D mode (~1350 cm$^{-1}$) only appears at the graphene edge; the mode around 1400 cm$^{-1}$ is an LED artifact. **F** Two-dimensional map of the D intensity as a function of double tip position (200 x280 nm$^2$ area with 10x7-pixel spatial resolution). The D band appears on the edge as determined from the topography image in A